\documentclass[conference]{IEEEtran}
\IEEEoverridecommandlockouts
\usepackage{cite}
\usepackage{amssymb}
\usepackage{algorithmic}
\usepackage{graphicx}
\usepackage{textcomp}
\usepackage{hyperref} 
\usepackage{xcolor}
\usepackage{comment}
\usepackage{multicol}
\def\BibTeX{{\rm B\kern-.05em{\sc i\kern-.025em b}\kern-.08em
    T\kern-.1667em\lower.7ex\hbox{E}\kern-.125emX}}
    
\usepackage{exscale,relsize}
\usepackage{graphicx}
\usepackage{enumerate}
\usepackage{array}
\usepackage{bigstrut}
\usepackage{extarrows}
    
\usepackage{amsmath}
\usepackage{amsfonts}

\usepackage{amsthm}

\usepackage[colorinlistoftodos,prependcaption,textsize=tiny]{todonotes}

\newtheorem{theorem}{Theorem}[section]
\newtheorem{problem}{Problem}[section]
\newtheorem{proposition}{Proposition}[section]
\newtheorem{remark}{Remark}

\newcommand{\6}{\mathbf}

\usepackage[scaled=0.3]{helvet}

\usepackage[lined,ruled,linesnumbered,noend]{algorithm2e}
\SetAlCapSkip{1em}

\usepackage{calligra}
\newcommand{\code}{\fontsize{8}{8}{\hspace{-1mm}{{\text{\calligra{C} }}}}}
\newcommand{\subcode}{\hspace{-1mm}{\fontsize{8}{8}{\text{\calligra{B} }}}}
\newcommand{\comb}{\hspace{-1mm}{\small{\text{\calligra{S} }}}}
\newcommand{\GL}{\text{GL}}

\usepackage{tikz}
\usepackage{pgfplots}

\definecolor{mygrey}{RGB}{192, 192, 192}

\newcommand{\smallgaussbinom}[2]{\left[\begin{smallmatrix}#1\\#2\end{smallmatrix}\right]_q}

\usepackage{scalerel,stackengine}
\stackMath

\newcommand\rhat[1]{%
\savestack{\tmpbox}{\stretchto{%
  \scaleto{%
    \scalerel*[\widthof{\ensuremath{#1}}]{\kern-.6pt\bigwedge\kern-.6pt}%
    {\rule[-\textheight/2]{1ex}{\textheight}}%WIDTH-LIMITED BIG WEDGE
  }{\textheight}% 
}{0.5ex}}%
\stackon[1pt]{#1}{\tmpbox}%
}

\newcommand\rcheck[1]{%
\savestack{\tmpbox}{\stretchto{%
  \scaleto{%
    \scalerel*[\widthof{\ensuremath{#1}}]{\kern-.6pt\rotatebox[origin = c]{180}{$\bigwedge$}\kern-.6pt}%
    {\rule[-\textheight/2]{1ex}{\textheight}}%WIDTH-LIMITED BIG WEDGE
  }{\textheight}% 
}{0.5ex}}%
\stackon[1pt]{#1}{\tmpbox}%
}
\begin{document}

\title{A Novel Attack to the Permuted Kernel Problem}

\author{Paolo Santini, Marco Baldi, Franco Chiaraluce\\

\IEEEauthorblockN{
%Dipartimento di Ingegneria dell'Informazione\\
Università Politecnica delle Marche\\ 
Ancona, Italy\\
\texttt{\{p.santini, m.baldi, f.chiaraluce\}@univpm.it}\\
} 
}
\maketitle

\begin{abstract}
The Permuted Kernel Problem (PKP) asks to find a permutation of a given vector belonging to the kernel of a given matrix.
The PKP is at the basis of PKP-DSS, a post-quantum signature scheme deriving from the identification scheme proposed by Shamir in 1989.  
The most efficient solver for PKP is due to a recent paper by Koussa et al. 
In this paper we propose an improvement of such an algorithm, which we achieve by considering an additional collision search step applied on kernel equations involving a small number of coordinates.
We study the conditions for such equations to exist from a coding theory perspective, and we describe how to efficiently find them with methods borrowed from coding theory, such as information set decoding.
We assess the complexity of the resulting algorithm and show that it outperforms previous approaches in several cases.
We also show that, taking the new solver into account, the security level of some instances of PKP-DSS turns out to be slightly overestimated.
\end{abstract}


\section{Introduction}
One of the oldest paradigms to achieve digital signatures consists in converting a Zero-Knowledge Identification (ZK-ID) scheme into a signature scheme through the Fiat-Shamir approach \cite{fiat1986prove}.
In a ZK-ID protocol a \textit{prover}, holding the secret key, proves their identity through an interactive procedure, by replying to random challenges provided by a \textit{verifier}.
Fiat-Shamir makes the protocol non interactive; in the resulting scheme, the signature corresponds to the transcript of the protocol, i.e, to the list of exchanged messages.
Usually, in a ZK-ID scheme, the key pair is generated by choosing a random instance of some hard problem: no trapdoor is involved and, consequently, the security guarantees are rather strong.

However, with a straightforward application of Fiat-Shamir, the resulting signatures are normally rather large.
For this reason, ZK-ID signatures have
received little attention for many years.
It seems, however, that this trend is changing, since several works describing modern ZK-ID signatures have recently appeared \cite{gueron2021designing, bettaieb2021zero, bidoux2022code, beullens2020sigma, beullens2019pkp, lessfm, joux}. 
These schemes make use of several optimizations, ranging from simulating a multiparty computation phase \cite{mpc} to using hash-based functions (e.g., PRNGs and tree structures), which can lead to compact signatures with essentially no impact on security.  
This renewed interest is also motivated by the fact that devising secure and efficient post-quantum digital signatures looks difficult, especially as concerns the possibility to achieve the advisable diversity with respect to the sole availability of schemes based on structured lattices \cite{thirdround,nistadditional}. 
ZK-ID signatures actually represent a promising and concrete avenue in this direction.

In 1989, Shamir proposed a ZK-ID protocol based on the Permuted Kernel Problem (PKP) \cite{shamir}.
This protocol is at the core of PKP-DSS \cite{beullens2019pkp}, a recently proposed signature scheme with competitive performance (e.g., public keys of 57 bytes, signatures of 20.5 kilobytes and constant time signing in 2.5 millions of cycles, for 128-bit security).
The PKP, which has been extensively studied along the years \cite{pkp_attack_1, pkp_attack_2, pkp_attack_3, pkp_attack_4, pkp_attack_5}, is an NP-hard problem \cite{garey1979computers} that asks to find the permutation of a given vector which belongs to the kernel of a given matrix.
The state-of-the-art PKP solver analyzed in the recent paper \cite{pkpsolve2019}, in a nutshell, works by first reducing the problem to a smaller instance of the same problem, which is then solved with a meet-in-the-middle search strategy.
The complexity of such an algorithm has been considered to recommend parameters for PKP-DSS.

In this paper we improve upon the state-of-the-art solver for the PKP.
Technically, our algorithm can be thought of as an improvement of the one in \cite{pkpsolve2019}, where we include a  filtering step to cut some of the elements in the initial lists.
To do this, we need to find kernel equations which bind a small number of coordinates. 
A similar idea has already been briefly discussed in \cite{pkp_attack_3, pkpsolve2019}; in both those works, however, the authors conclude that such equations are extremely hard to find and that, in practice, cannot be exploited.
We adopt a coding theory perspective and show that, instead, useful equations of this type can be efficiently found by exploiting Information Set Decoding (ISD) algorithms. 
The resulting solver runs in a time which is lower than that of \cite{pkpsolve2019} and can attack some of the instances recommended for PKP-DSS (namely, those for $128$ and $192$ bits of security) with a smaller complexity than that claimed in \cite{beullens2019pkp}.
The performance of the proposed algorithm has been tested with a proof-of-concept software implementation, which is publicly available\footnote{\url{https://github.com/secomms/pkpattack/}}.

The paper is organized as follows.
In Section \ref{sec:preliminaries} we settle the notation we use throughout the paper and provide some basic notions about linear codes.
In Section \ref{sec:state} we briefly recall the definition of PKP and the algorithm in \cite{pkpsolve2019}.
In Section \ref{sec:subcodes} we describe how to find kernel equations with the desired properties.
In Section \ref{sec:attack} we describe and analyze the new PKP solver.
In Section \ref{sec:conclusions} we draw some conclusive remarks.

\section{Notation and preliminaries} \label{sec:preliminaries}
In this section we define the notation we use throughout the paper and recall some basic notions about linear codes.

\subsection{Notation}

We use $\mathbb F_q$ to denote the finite field with $q$ elements.
Bold lowercase (resp., uppercase) letters indicate vectors (resp., matrices).
Given $\6a$ (resp., $\6A$), $a_i$ (resp., $a_{i,j}$) denotes the entry in position $i$ (resp., the entry in the $i$-th row and $j$-th column).
$\GL_{m,n}$ is the set of $m\times n$ matrices over $\mathbb F_q$ with full rank $\min\{m,n\}$. 
The identity matrix of size $n$ is indicated as $\6I_n$, while $\60$ denotes the all-zero vector.
%By support  of a matrix corresponds to the set of indexes pointing at non-null columns; analogous notation is used for vectors.
Given a set $A$, $|A|$ denotes its cardinality (i.e., the number of elements) and $a\xleftarrow{\$}A$ means that $a$ is picked uniformly at random over $A$.
Given a matrix $\6A$ and a set $J$, $\6A_J$ is the matrix formed by the columns of $\6A$ that are indexed by $J$; analogous notation is used for vectors.
We denote by ${\sf{RREF}}(\6A,J)$ the algorithm that outputs $\6A_J^{-1}\6A$ if $\6A_J$ is square and non singular, otherwise returns a failure.
We use $S_n$ to denote the group of length-$n$ permutations.
Given $\6a = (a_1,\cdots,a_n)$ and $\pi\in S_n$, we write $\pi(\6a) = \big(a_{\pi(1)},\cdots,a_{\pi(n)}\big)$.
Given $\6a$, $\6b$, we define
$\6a\cap \6b$ as the set of entries which appear in both $\6a$ and $\6b$.
%\item[-] $\6a\setminus \6b$ as the set of entries that are in $\6a$ but not in $\6b$.
%\item[-] $\6a\subseteq \6b$ if the set formed by the entries of $\6a$ is contained in the set formed by the entries of $\6b$.\end{itemize}
For a vector $\6a\in\mathbb F_q^n$ with no repeated entries, we define $\comb_{\ell}(\6a)$ as the set of length-$\ell$ vectors with entries picked from those of $\6a$.
Notice that $|\hspace{0.8mm}\comb_\ell(\6c)| = \frac{n!}{(n-\ell)!}$.

\subsection{Linear codes}

A linear code $\code \subseteq \mathbb F_q^n$ with dimension $k$ and redundancy $r = n-k$ is a linear $k$-dimensional subspace of $\mathbb F_q^n$.
Any code admits two equivalent representations: a \textit{generator matrix}, that is, any $\6G\in\GL_{k,n}$ such that $\code = \{\6u\6G\mid \6u\in\mathbb F_q^k\}$, or a
\textit{parity-check matrix}, that is, any $\6H\in\GL_{r,n}$ such that $\code = \{\6c\in\mathbb F_q^n\mid \6c\6H^\top = \60\}$ (where $\top$ denotes transposition).
Given $\6x\in\mathbb F_q^n$, its syndrome is $\6s = \6x\6H^\top$.
The dual of $\code$, which we denote by $\code^\bot$, is the space generated by $\6H$. %\textit{i.e.}, the code having $\6H$ and $\6G$ as generator and parity-check matrices, respectively.
For any codeword $\6c\in\code$ and any  $\6b\in\code^\bot$, we have $\6c\6b^\top = 0$.
By support of a code we mean the set of indexes $i$ such that there is at least one codeword $\6c$ with $c_i\neq 0$. 
A subcode $\subcode\subseteq \code$, with dimension $k'$, is a $k'$-dimensional linear subspace of $\code$.
The number of such subcodes is counted by $\smallgaussbinom{k}{k'} = \prod_{i = 0}^{k'-1} \frac{1-q^{k-i}}{1-q^{i+1}}$.

\section{The Permuted Kernel Problem\label{sec:state}}

The Permuted Kernel Problem (PKP) reads as follows.
\begin{problem}\label{prob:pkp}
\textbf{Permuted Kernel Problem (PKP)}
\\Given $\6A\in\mathbb F_q^{m\times n}$ with $1\leq m< n$ and $\6c\in\mathbb F_q^{n}$, find $\pi \in S_n$ such that $\pi(\6c)\6A^\top = \60$.
\end{problem}
The problem is notably known to be NP-hard, via reduction from the Subset Sum Problem (SSP) \cite{garey1979computers}.
In the following sections we briefly recall the features of the hardest PKP instances and  recall the algorithm in \cite{pkpsolve2019}, which is deemed as the currently known best solver for PKP.

\begin{remark} 
The PKP can be  equivalently formulated as a codeword finding problem.
In fact, Problem \ref{prob:pkp} asks to find a codeword $\widetilde{\6c}\in\code$, where $\code$ is the code having $\6A$ as parity-check matrix, such that $\widetilde{\6c}\in \comb_n(\6c)$.
\end{remark}

\subsection{Considerations for practical hardness}

As in all previous works \cite{pkp_attack_1, pkp_attack_2, pkp_attack_3, pkp_attack_4, pkp_attack_5, pkpsolve2019}, we study the PKP under the conditions leading to the hardest instances.
Namely, we consider $\6A$ such that $\mathrm{rank}(\6A) = m$, $\6c$ with all distinct entries and consider parameters $q,n,m$ so that, on average, the problem has exactly one solution.
To this end, we assume that the PKP instance is generated by first picking $\6A\xleftarrow{\$}\GL_{m,n}$ and then by choosing a random vector $\widetilde{\6c}\in\mathbb F_q^n$ with distinct entries and such that $\widetilde{\6c}\6A^\top = \60$. Then, we set $\6c =  \pi(\widetilde{\6c})$, with  $\pi\xleftarrow{\$}S_n$.
%since the entries of $\6c$ are all distinct, we have that the orbit of $\6c$ under the action of permutations corresponds to $\comb_n(\6c)$. 
%Hence, trivially, solving PKP consists in finding $\widetilde{\6c}\in\comb_n(\6c)$ such that $\widetilde{\6c}\6A^\top = \60$.
%Indeed, for any such $\widetilde{\6c}$, we always have a permutation $\sigma\in S_n$ for which $\sigma(\widetilde{\6c}) = \6c$.
%Note that the number of solutions is equal to the number of vectors in $\comb_n(\6c)$ that are also in the kernel of $\6A$.
Since $\6A$ and $\widetilde{\6c}$ are picked at random, on average we expect to have
$\frac{|\hspace{1mm}\comb_n(\6c)|}{q^m} = \frac{n!}{q^m}$ solutions.
Consequently, we consider $q,n,m$ such that $n!q^{-m}<1$.

Basically any solver for the PKP considers that it is always possible to craft additional constraints binding the entries of $\widetilde{\6c}$.
%Indeed, let us consider a function that i) takes as input the coordinates of $\6c$ and ii) is invariant under the action of permutations (\textit{i.e.}, the function values for inputs from $\comb_{n}(\6c)$ are all the same).
Namely, we can exploit any relation of the form
\begin{equation}
\label{eq:new_relation}
 \sum_{i = 1}^n \tilde c_i^{\hspace{1mm}u} = \sum_{i = 1}^n c_i^{\hspace{1mm}u},  \hspace{2mm}u\in \{1, \cdots, q-1\}.
\end{equation}
However, for $u\geq 2$ the above expression is not linear in the unknowns $\tilde c_i$, so that only the case of $u = 1$ is employed.
%so that the corresponding extra constraint may not be useful in the search for the solution $\widetilde{\6c}$.

Taking into account all the previous considerations, the PKP formulation in Problem \ref{prob:pkp} can be slightly modified.
Indeed, let $\6H = \begin{pmatrix}
\begin{matrix}\6A\end{matrix}\\
\begin{matrix}1 & \cdots & 1\end{matrix}
\end{pmatrix}\in\mathbb F_q^{r\times n}$, with $r = m+1$.
Then, solving the PKP corresponds to finding $\widetilde{\6c}\in \comb_n(\6c)$ such that 
\begin{align}
\label{eq:new_pkp}
\widetilde{\6c}\6H^\top =  \big(0,\cdots,0,\sum_{i = 1}^nc_i\big) = \6s.
\end{align}
With overwhelming probability (approximately $1-q^{-m}$), the all-ones vector is not a linear combination of the rows of $\6A$, so that we can safely assume that $\6H$ has full rank $r$.

Finally, we consider that to solve the PKP we can restrict our attention to a subset of the entries of $\widetilde{\6c}$.
Indeed, for any $\6B\in\GL_{\ell,r}$ with $\ell\leq r$, it must be
\begin{equation}
\label{eq:subcode_key_equation}
\widetilde{\6c}(\6B\6H)^\top = \widetilde{\6c}\widetilde{\6H}^\top = \6s\6B^\top = \widetilde{\6s}.
\end{equation}

Let $J\subset \{1,\ldots,n\}$ of size $n-r$ such that $\6H_{\{1,\cdots,n\}\setminus J}$ is non singular, and
$\6B = {\6H^{-1}_{\{1,\cdots,n\}\setminus J}}$.  
Then, $\widetilde{\6H} = \6B\6H = {\sf{RREF}}(\6H,\{1,\cdots,n\}\setminus J)$, from which
\begin{equation}
\label{eq:key_relation}
\tilde c_{i_u} = \tilde s_i - \sum_{j\in J}\tilde c_j \tilde h_{j,u},\hspace{2mm}\{i_1,\cdots,i_r\} = \{1,\cdots,n\}\setminus J.  
\end{equation}
Hence, it is enough to find the entries of $\widetilde{\6c}$ in the positions indexed by $J$ to retrieve the whole solution $\widetilde{\6c}$.

\begin{remark}
Adopting again a coding theory formulation, one can see the PKP as a syndrome decoding problem: 
given a parity-check matrix $\6H$ and a syndrome $\6s$ as in \eqref{eq:new_pkp}, find a vector $\widetilde{\6c}\in\comb_n(\6c)$ whose syndrome is $\6s$.
\end{remark}

\subsection{State-of-the-art solver for PKP}

The currently known best solver for the PKP is Algorithm 1 in \cite{pkpsolve2019}.
The algorithm works with three parameters $\ell, \ell_1,\ell_2\in\mathbb N$, such that $1\leq \ell \leq r$, $\ell_1,\ell_2\geq 1$ and $\ell_1+\ell_2 = n-r+\ell$.
The procedure is initialized by choosing a matrix $\6B\in\GL_{\ell,r}$ so that $\widetilde{\6H} = \6B\6H$ has support size $n-r+\ell$.
To do this, we first compute $\6H' = {\sf{RREF}}(\6H,\{n-r+1,\cdots,n\})$  and then sets $\widetilde{\6H}$ as the sub-matrix formed by the entries of $\6H'$ in the first $\ell$ rows and the columns in positions $\{1,\cdots,n-r+\ell\}$.
The same transformation is applied to $\6s$, obtaining $\widetilde{\6s} = \6s\6B^\top\in\mathbb F_q^\ell$.
Then, we partition $\widetilde{\6H}$ as $(\widetilde{\6H}_1, \widetilde{\6H}_2)$, where $\widetilde{\6H}_1\in\mathbb F_q^{\ell\times \ell_1}$ and $\widetilde{\6H}_2\in\mathbb F_q^{\ell\times \ell_2}$, and construct two lists
$$\mathcal L_1 = \left\{\left.(\6x,\6x\widetilde{\6H}_1^\top)\right|\6x\in\comb_{\ell_1}(\6c)\right\},$$
$$\mathcal L_2 = \left\{\left.(\6y,\widetilde{\6s}-\6y\widetilde{\6H}_2^\top)\right|\6y\in\comb_{\ell_2}(\6c)\right\}.$$
Let $\mathcal L = \mathcal L_1 \bowtie \mathcal L_2$, where $\bowtie$ is computed as follows:
\begin{enumerate}
\item use an efficient search algorithm (e.g., permutation plus binary search) to find collisions, i.e., pairs $(\6x,\6t)\in\mathcal L_1$ and $(\6y,\6v)\in\mathcal L_2$ such that $\6t = \6v$;
\item keep only the collisions for which $\6x\cap \6y = \varnothing$.
\end{enumerate}
By construction, $\mathcal L = \left\{(\6x,\6y)\in\comb_{\ell_1+\ell_2}(\6c)\mid (\6x,\6y)\widetilde{\6H}^\top = \widetilde{\6s}\right\}$.
%The found collisions are filtered, so that we keep only those for which $\6x\cap \6y = \varnothing$.
%The collisions are used to build vectors $(\6x,\6y)\in\comb_{\ell_1+\ell_2}(\6c)$ which, by construction, are such that $(\6x,\6y)\6H'^\top = \6s'$, 
Then, we find $J$ of size $n-r$ so that $J\subseteq\{1,\cdots,n-r+\ell\}$ and $\6H_{\{1,\cdots,n\}\setminus J}$ is non singular, compute $\widetilde{\6H} = {\sf{RREF}}(\6H,\{1,\cdots,n\}\setminus J)$ and use  \eqref{eq:key_relation} to test each element in $\mathcal L$.
Namely, for each $\6p\in\mathcal L$, we use the entries of $\6p_J$ as $\widetilde{\6c}_J$ and see if the resulting $\widetilde{\6c}$ belongs to $\comb_{n}(\6c)$. 

\begin{comment}
\paragraph{\textbf{Time complexity}}

According to \cite{pkpsolve2019}, the time complexity of the algorithm can be easily assessed with the following considerations:
\begin{itemize}
\item[-] to construct the lists, we use a time complexity given by $$|\mathcal L_1|+|\mathcal L_2| = \frac{n!}{(n-\ell_1)!}+\frac{n!}{(n-\ell_2)!};$$
\item[-] to merge 
\item[-] the cost of the search phase in the computation of $\mathcal L_1\bowtie \mathcal L_2$ can be estimated as 
$\max\left\{|\mathcal L_1|, |\mathcal L_2|\right\}$;
\item $\frac{|\mathcal L_1|\cdot|\mathcal L_2|}{q^\ell}$
$ = \frac{(n!)^2}{q^\ell (n-\ell_1)!(n-\ell_2)!};$
\item[-] after the filtering, we remain with a number of candidates given by
$$\frac{|\hspace{0.5mm}\comb_{n-r+\ell}(\6c)|}{q^\ell} = \frac{n!}{q^\ell(n-r+\ell)!}.$$
Notice that this number is, obviously, always smaller than the number of collisions.
\end{itemize}
\end{comment}
According to \cite{pkpsolve2019}, the time complexity of the algorithm is given by
\begin{equation}
\label{eq:time_state}
T(\ell_1,\ell_2) = \frac{n!}{(n-\ell_1)!}+\frac{n!}{(n-\ell_2)!} + \frac{(n!)^2q^{n-r-\ell_1-\ell_2}}{(n-\ell_1)!(n-\ell_2)!}.
\end{equation}
%In practice, the algorithm is optimized when the number of collisions is almost identical to the size of the initial lists.

\section{Finding subcodes with small support}
\label{sec:subcodes}
Next we show that, differently from the claims in \cite{pkp_attack_3, beullens2019pkp}, we can efficiently find kernel equations which involve a small number of coordinates.
We first substantiate the existence of such equations with coding theory arguments, and then describe how to efficiently find them.
%new way to solve the problem.
%Basically, our proposed algorithm consists in the same approach of \cite{pkpsolve2019}, with the only difference that we consider an additional phase in which we filter elements from the initial lists, exploiting subcodes of $\code^\bot$ having small support.

\subsection{Number of subcodes with small support}

As shown above, we can see the matrix $\6H$ of a given PKP instance as the parity-check matrix of some linear code $\code$ with redundancy $r$. 
The space generated by the rows of $\6H$ corresponds to $\code^\bot$, and 
%Since $\6A$ is picked at random among $\GL_{m,n}$, we have that also $\code$ is picked at random among the codes with fixed redundancy.
%A kernel equation involving $w$ coordinates corresponds to a codeword $\6a\in\code^\bot$ with Hamming weight $w$.
a set of $d\leq r$ independent equations from this space, involving $w$ coordinates, is a basis for a subcode $\subcode\subseteq\code^\bot$ with dimension $d$ and support size $w$.
For a random code, the number of such subcodes can be estimated as follows.
\begin{theorem}\label{prop:subcodes}
For a code $\code\subseteq \mathbb F_q^n$, we define $\mathcal A_{w,d}(\code\hspace{0.8mm})$ as the set of subcodes of $\code$ with dimension $d$ and support size $w$. 
Let $N_{w,d}$ be the average value of $|\mathcal A_{w,d}(\code\hspace{0.8mm})|$, when $\code$ is picked at random among all codes with dimension $k$.
Then $\rcheck N_{w,d}\leq N_{w,d}\leq \rhat N_{w,d}$, with
$$\rcheck N_{w,d}= \binom{n}{w}(q^d-1)^{w-d}\frac{\smallgaussbinom{k}{d}}{\smallgaussbinom{n}{d}},$$
$$\rhat N_{w,d}=\binom{n}{w}\frac{(q^d-1)^w}{\prod_{i = 0}^{d-1}(q^d-q^i)}\frac{\smallgaussbinom{k}{d}}{\smallgaussbinom{n}{d}}.$$
\end{theorem}
\begin{IEEEproof}
Let $\mathcal U_k$ be the set of all linear codes over $\mathbb F_q$ with length $n$ and dimension $k$.
We observe that
\begin{align*}
%N_{w,d} & = \frac{\sum_{\code \in \mathcal U_k}\sum_{\hspace{1mm}\subcode\in \mathcal A_{w,d}(\mathbb F_q^n)} p(\subcode,\code\hspace{0.8mm})}{|\mathcal U_k|}\\\nonumber
N_{w,d} & = \frac{\sum_{\code \in \mathcal U_k}|\mathcal A_{w,d}(\code\hspace{0.5mm})|}{|\mathcal U_k|}\\\nonumber
& = \frac{\sum_{\code \in \mathcal U_k}\sum_{\hspace{1mm}\subcode\in \mathcal A_{w,d}(\mathbb F_q^n)} p(\hspace{0.5mm}\subcode,\code\hspace{0.5mm})}{\smallgaussbinom{n}{k}},
\end{align*}
where $p(\hspace{0.5mm}\subcode,\code\hspace{0.5mm}) = 1$ if $\subcode\subseteq\code$, and $0$ otherwise.
With a simple rewriting, we obtain 
\begin{align*}
N_{w,d}  & = 
\frac{\sum_{\subcode \in \mathcal A_{w,d}(\mathbb F_q^n)}\sum_{\code \in \mathcal U_k} p(\hspace{0.5mm}\subcode,\code\hspace{0.5mm})}{\smallgaussbinom{n}{k}}\\\nonumber
& = \frac{\sum_{\subcode \in \mathcal A_{w,d}(\mathbb F_q^n)}\smallgaussbinom{n-d}{k-d}}{\smallgaussbinom{n}{k}},
\end{align*}
where the r.h.s. term is justified by the observation that $\sum_{\code \in \mathcal U_k} p(\hspace{0.5mm}\subcode,\code\hspace{0.5mm})$ is equal to the number of $k$-dimensional codes having $\subcode$ as a subcode; this quantity is given by $\smallgaussbinom{n-d}{k-d}$ (that is, the number of $(k-d)$-dimensional subspaces of $\mathbb F_q^n\setminus \hspace{0.8mm}\subcode$, which has dimension $n-d$).
With simple algebra, we find that $\smallgaussbinom{n-d}{k-d}/\smallgaussbinom{n}{k} = \smallgaussbinom{k}{d}/\smallgaussbinom{n}{d}$.
So, we further obtain
$$N_{w,d} = |\mathcal A_{w,d}(\mathbb F_q^n)|\frac{\smallgaussbinom{k}{d}}{\smallgaussbinom{n}{d}}.$$
If the support of a subcode is $J$, then any of its generator matrices must be such that the columns indexed by $J$ are non-null.
For a fixed $J$, the number of such matrices is $(q^d-1)^w$: to consider that any code has multiple generator matrices, we divide this quantity by the number of changes of basis, that is, $\prod_{i = 0}^{d-1}(q^d-q^i)$.
This way we obtain an upper bound on the size of $|\mathcal A_{w,d}(\mathbb F_q^n)|$: indeed, some of the matrices we are considering may have rank $<d$.
Considering that we have $\binom{n}{w}$ choices for $J$, we obtain an upper bound since
$$|\mathcal A_{w,d}(\mathbb F_q^n)|\leq \binom{n}{w}\frac{(q^d-1)^w}{\prod_{i = 0}^{d-1}(q^d-q^i)}.$$
To prove the lower bound, we fix again a set $J$ and, among all the matrices with support $J$, consider only those for which the leftmost $d\times d$ submatrix is the identity matrix.
This way, we avoid multiple counting of the same code: any two matrices $(\6I_d, \6V)$ and $(\6I_d, \6V')$ (restricting to the columns indexed by $J$) such that $\6V\neq \6V'$ will generate different codes.
Note that a matrix $(\6I_d, \6V)$ can generate a code with support size $w$ if and only if $\6V\in\mathbb F_q^{d\times(w-d)}$ has no null column: the number of such matrices is $(q^d-1)^{w-d}$.
This way we obtain a lower bound, since there exist also codes that do not admit a generator matrix in the form $(\6I_d, \6V)$.
Considering again the number of choices for $J$, we set a lower bound as
$$|\mathcal A_{w,d}(\mathbb F_q^n)|\geq \binom{n}{w}(q^d-1)^{w-d}.$$

\vspace{-1.7em}
\end{IEEEproof}

\begin{comment}
\begin{remark}
The bound expressed in the above proposition is expected to be tight for small values of $w$.
Indeed, analogously to the concept of minimum distance, subcodes of a random code are expected to have a minimum support size which cannot be lower than some threshold value.
As a rule of thumb, we can estimate the minimum support size as the lowest $w^*$ such that $N_{w^*,d}>1$.
If we choose $w$ equal to $w^*$ (or only slightly larger), then we expect the bound in Proposition \ref{prop:subcodes} to be tight. 
\end{remark}
\end{comment}
\begin{remark}
When $d = 1$, a subcode corresponds to the orbit of a codeword under scalar multiplication by the elements of $\mathbb F_q$. 
The bounds in Theorem \ref{prop:subcodes} coincide, so that  
$$N_{w,1} = \binom{n}{w}(q-1)^{w-1}\frac{q^k-1}{q^n-1}\approx \binom{n}{w}(q-1)^{w-1}q^{k-n}.$$
\end{remark}
\begin{comment}
To this end, we consider the following proposition.
\begin{proposition}\label{prop:codewords}
Let $\6H\xleftarrow{\$}\GL_{r,n}$, and let $\code$ be the code generated by $\6H$.
Let $\mathcal A_{w,1}$ be the set of all subcodes of $\code$ with support size $w$ and dimension $1$.
Then the average value of $\left|\mathcal A_{w,1}\right|$ is $N_w$, where
$$N_w = \binom{n}{w}(q-1)^{w-1}q^{k-n}.$$
\end{proposition}
\begin{IEEEproof}
Let $J\subseteq \{1,\cdots,n\}$ be a set with size $w$.
For $\6a \in\mathbb F_q^n$ with weight $w$, we define $\subcode(\6a) = \left\{b\6a\mid b\in \mathbb F_q\right\}$.
To fully represent each $\subcode(\6a)$, we can use the non null element that comes first in lexicograph order, that is, the one having $1$ as the first non null entry.
This way, for each $J$, we count $(q-1)^{w-1}$ distinct subcodes $\subcode(\6a)$.
Since $\code$ is linear, we have $\subcode(\6a)\subseteq \code$ if and only if $\6a\in\code$.
Since $\6H$ is picked at random, we can assume that $\6a$ is a codeword with probability $q^{k-n}$.
Hence, for each choice for $J$, we expect to have $(q-1)^{w-1}q^{k-n}$ codewords with weight $w$ and support $J$.
Multiplying this quantity by the number of choices for $J$ (that is, $\binom{n}{w}$), we prove the thesis.
\end{IEEEproof}
\end{comment}

\subsection{Using ISD to find subcodes with small support}

The result in Theorem \ref{prop:subcodes} can be used to set values for $w$ and $d$ such that, given a random code $\code$, $\mathcal A_{w,d}(\hspace{0.5mm}\code\hspace{0.5mm})$ is non empty with high probability.
As a rule of thumb, we consider that whenever $\rcheck{N}_{w,d}>1$, the code contains at least one subcode with the desired properties.
When $w$ is much smaller than $n$, then such subcodes can be efficiently found using ISD algorithms.
If $d = 1$, finding subcodes with small support is equivalent to find codewords with small Hamming weight: we consider the algorithm in \cite{PetersFq} and denote its time complexity as $T^{(1)}_{\scaleto{\mathcal{ISD}}{4pt}}(n,k,w)$.
%as an example of efficient ISD working over $\mathbb F_q$, 
When $d>1$, we can apply minor tweaks to ISD algorithms and use them to find subcodes.
To the best of our knowledge, this idea has been considered only in \cite{beullens}, for the case of 2-dimensional codes and adapting Prange's simple ISD \cite{prange}.
We consider a generalization of this method, where subcodes can have any dimension $d$; the corresponding procedure is detailed in Algorithm \ref{isd}.

\begin{algorithm}[ht]\label{isd}
\small
\KwIn{generator matrix $\6G\in\GL_{k,n}$ for $\code$, $w,d\in\mathbb N$}
\KwOut{failure, or generator matrix for $\subcode\subseteq \code$ with dimension $d$ and support size $w$}
%\vspace{1em}
\SetAlgoNoLine
$\sigma\xleftarrow{\$}S_n$\;
\eIf{${\sf{RREF}}\big(\sigma(\6G),\{1,\cdots,k\}\big)$ \emph{fails}}{
Report failure\;
}{
$(\6I_d, \6V)\gets {{\sf{RREF}}\big(\sigma(\6G),\{1,\cdots,k\}\big)}$
}
\vspace{0.5em}
\SetAlgoVlined
\For{$U\subseteq\{1,\cdots,k\}$ \emph{with size $d$}}{
$\6B\gets$ matrix formed by rows of $\6V$ indexed by $U$\;
\SetAlgoNoLine
\If{$\6B$ \emph{has support size $w-d$}}{
Return $\sigma^{-1}\big((\6I_d, \6B)\big)$}
}
Report failure\;
 \caption{One  iteration of ISD for $d>1$}
\end{algorithm}
For the algorithm to work, it must be $w\leq n+d-k$.
The probability that the computation of ${\sf{RREF}}$ does not fail can be estimated as $\frac{\prod_{i = 0}^{d-1}q^d-q^{i}}{q^{d^2}}$ and, for large $q$, it can be assumed to be equal to $1$.
Let $\subcode\subseteq \mathcal A_{w,d}(\hspace{0.5mm}\code\hspace{0.5mm})$; then, the probability that one iteration finds $\subcode$ is given by 
$p(n,k,d,w) = \frac{\binom{w}{d}\binom{n-w}{k-d}}{\binom{n}{k}}$.
When we have $|\mathcal A_{w,d}(\hspace{0.5mm}\code\hspace{0.5mm})|$ subcodes and we are simply interested in finding one of them, the success probability can be estimated as $1-\left(1-p(n,k,d,w)\right)^{|\mathcal A_{w,d}(\hspace{0.5mm}\code\hspace{0.5mm})|}$.
By using the lower bound in Theorem \ref{prop:subcodes}, we conservatively set this probability as
$1-\left(1-p(n,k,d,w)\right)^{\rcheck N_{w,d}}$.
Computing ${\sf{RREF}}$ comes with a broad cost of $O(k^3)$, while the number of sets $U$ that are tested is $\binom{k}{d}$. 
Consequently, we assess the cost of finding a subcode of $\mathcal A_{w,d}(\hspace{0.5mm}\code\hspace{0.5mm})$ as  
\begin{equation}
\label{eq:isd}
T^{(d)}_{\scaleto{\mathcal{ISD}}{4pt}}(n,k,w) = O\left( \frac{k^3+\binom{k}{d}}{1-\left(1-p(n,k,d,w)\right)^{\rcheck N_{w,d}}}\right).    
\end{equation}

\section{New PKP solver}
\label{sec:attack}
\begin{figure}[h!]
\centering
\resizebox{0.95\columnwidth}{!}{
\begin{tikzpicture}

\draw[line width=1mm, mygrey] (4, 5) to (9,0);
\fill[mygrey] (0, 0) rectangle (4, 5);

%\fill[mygrey] (5, 0) rectangle (8, 4);
%\fill[mygrey] (1, 3) rectangle (2.5, 4);

%\draw (2.5, 4) to (2.5,0);
%\draw (0, 1.5) to (5,1.5);

\draw (0, 0) rectangle (9, 5);
%\draw (0, 1) rectangle (7, 4);

%\draw[line width=1mm, mygrey] (1.15, 2.9) to (3.85,0.1);
%\node (a) at (0.5, 3.5) {\huge$\6I_{k_p}$};

%\node (a) at (1.75, 3.5) {\huge$\6A$};
%\node (a) at (0.5, 2.25) {\huge$\60$};
%\node (a) at (0.5, 0.75) {\huge$\60$};
%\node (a) at (1.75, 0.75) {\huge$\60$};

%%matrix
%\node (a) at (0.15, 3.75) {\Large 1};
%\node (a) at (0.85, 3.25) {\Large 1};
%\node (a) at (0.5, 3.6) {$\ddots$};

%\node (a) at (1.15, 2.75) {\Large 1};
%\node (a) at (2.35, 1.75) {\Large 1};
%\node (a) at (2.65, 1.25) {\Large 1};
%\node (a) at (3.85, 0.25) {\Large 1};
%\node (a) at (1.7, 2.6) {\Large $\ddots$};
%\node (a) at (2.5, 1.7) {\Large $\ddots$};
%\node (a) at (3.3, 0.7) {\Large $\ddots$};

%\node (a) at (1.4, 2.6) {\Large $\cdot$};
%\node (a) at (1.5, 2.5) {\Large $\cdot$};
%\node (a) at (1.6, 2.4) {\Large $\cdot$};

%\node (a) at (1.8, 2.2) {\Large $\cdot$};
%\node (a) at (1.9, 2.1) {\Large $\cdot$};
%\node (a) at (2, 2) {\Large $\cdot$};

%\node (a) at (2.4, 1.6) {\Large $\cdot$};
%\node (a) at (2.5, 1.5) {\Large $\cdot$};
%\node (a) at (2.6, 1.4) {\Large $\cdot$};

%\node (a) at (3-0.05, 1) {\Large $\cdot$};
%\node (a) at (3.1-0.05, 0.9) {\Large $\cdot$};
%\node (a) at (3.2-0.05, 0.8) {\Large $\cdot$};

%\draw (0, 3) to (5,3);
%\draw (1, 0) to (1,4);

\draw[<->] (0, -0.1) to node[anchor = north] {\large $n-r+\ell-w$} (2.68, -0.1);

\draw[<->] (2.72, -0.1) to node[anchor = north] {\large $w$} (7.8, -0.1);

%\draw[<->] (0, -0.7) to node[anchor = north] {\large $n-r+\ell$} (7.8, -0.7);

%\draw[<->] (3.52, 5.1) to node[anchor = south] {\large $\left\lceil \frac{n-r+\ell}{2} \right\rceil$} (7, 5.1);

%\draw[<->] (0, 6) to node[anchor = south] {\large $n-r+\ell$} (7, 6);

\draw[<->] (-0.1, 3.98) to node[anchor = east] {\large $\ell-d$} (-0.1, 1.2);

\draw[<->] (-0.1, 5) to node[anchor = east] {\large $d$} (-0.1, 4.02);

\draw (0, 4) to (7.8, 4);
\draw (2.7, 1.2) to (2.7, 4);
\draw (7.8, 1.2) to (7.8, 4);
\draw (0, 1.2) to (7.8, 1.2);

%\draw [decorate, decoration={brace,mirror, amplitude=10pt},xshift=0pt,yshift=-0.5pt] (0, 1.18) -- (2.7, 1.18) node [black,midway,yshift=-0.7cm] 
%{\large $\widetilde{\6H}_1$};

\node(a) at (1.35, 2.6) {\Large $\widetilde{\6H}_{L_1}$};
\node(a) at (5.25, 2.6) {\Large $\widetilde{\6H}_{L_2}$};

%\draw [decorate, decoration={brace,mirror, amplitude=10pt},xshift=0pt,yshift=-0.5pt] (2.7, 1.18) -- (7.8, 1.18) node [black,midway,yshift=-0.7cm] {\large $\widetilde{\6H}_2$};

\node (a) at (-1.7, 2.4) {\Large $=$};
\node (a) at (-3, 2.4) {\Large $\6M\sigma(\6H)$};

\fill[mygrey] (2.7, -3) rectangle (7.8, -2);
\draw (0, -3) rectangle (9, -2);
\draw (5, -2) to (5, -3);
\draw (7.8, -2) to (7.8, -3);
\draw (2.7, -2) to (2.7, -3);

%\draw[<->] (0, -1.9) to node[anchor = south] {\large $w$} (3, -1.9);

\node (a) at (-1.3, -2.5) {\Large $=$};
\node (a) at (-2.7, -2.5) {\Large $\sigma(\rhat{\6H})$};

\node (a) at (3.85, -2.5) {\Large $\6Z_{K_1}$};
\node (a) at (6.4, -2.5) {\Large $\6Z_{K_2}$};

\draw[<->] (-0.1, -2) to node[anchor = east] {\large $d$} (-0.1, -3);
\draw[<->] (0, -3.8) to node[anchor = north] {\large $n-r+\ell$} (7.8, -3.8);
\draw[<->] (0, -3.1) to node[anchor = north] {\large $n-r+\ell-w$} (2.68, -3.1);
\draw[<->] (2.72, -3.1) to node[anchor = north] {\large $\textcolor{white}{\ell}w_1\textcolor{white}{\ell}$} (4.98, -3.1);
\draw[<->] (5.02, -3.1) to node[anchor = north] {\large $\textcolor{white}{\ell}w_2\textcolor{white}{\ell}$} (7.8, -3.1);

%%%%%%%%%%%%%%%%%%%%%%%%%%%%%%%%%%%%%%%%%

\draw [decorate, decoration={brace, amplitude=10pt},xshift=0pt,yshift=-0.5pt](2.7, -1.9) -- (4.98, -1.9) node [black,midway,yshift=0.6cm] {\Large $\mathcal K_1$};
\draw [decorate, decoration={brace, amplitude=10pt},xshift=0pt,yshift=-0.5pt](5.02, -1.9) -- (7.8, -1.9) node [black,midway,yshift=0.6cm] {\Large $\mathcal K_2$};

\draw [decorate, decoration={brace, amplitude=10pt},xshift=0pt,yshift=-0.5pt](0, 5.1) -- (2.68, 5.1) node [black,midway,yshift=0.6cm] {\Large $\mathcal L_1$};
\draw [decorate, decoration={brace, amplitude=10pt},xshift=0pt,yshift=-0.5pt](2.72, 5.1) -- (7.8, 5.1) node [black,midway,yshift=0.6cm] {\Large $\mathcal L_2$};

\end{tikzpicture}
}
\caption{Representation of the operations of Algorithm \ref{alg:novel}.}
\label{fig:H}
\end{figure}
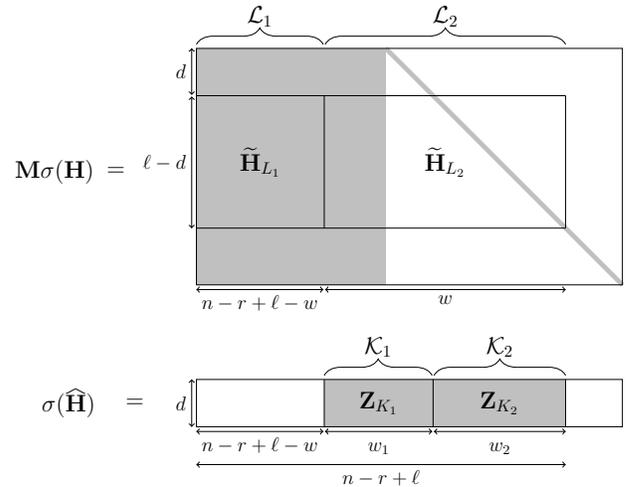

In this section we describe and analyze the algorithm we propose to solve the PKP. 
The method we propose is described in Algorithm \ref{alg:novel} and represented in Figure \ref{fig:H}.

\begin{algorithm}[ht]
\small
\KwData{$w,w_1,w_2,d,\ell\in\mathbb N$, such that $w\leq n$, $w = w_1 + w_2$, $d\leq r$, $\ell\leq n-r$.}
\KwIn{$\6H\in\GL_{r,n}$, $\6s\in\mathbb F_q^r$, $\6c\in\mathbb F_q^n$} 
\KwOut{$\widetilde{\6c}\in \comb_n(\6c)$ such that $\widetilde{\6c}\6H^\top = \6s$}
%%%%%%%%%%%%%%%%%%%%
\vspace{1em}
Use ISD to find $\rhat{\6H}$, generator matrix of $\subcode\subseteq\code^\bot$, with dimension $d$ and support size $w$\;
Compute $\6S\in\GL_{d,r}$ such that $\rhat{\6H} = \6S\6H$, $\sigma \in S_n$ such that $\mathrm{Supp}\big(\sigma(\rhat{\6H})\big) = \{n-r+\ell-w+1,\cdots,n-r+\ell\}$\;
$\rhat{\6s}\gets \6s\6S^\top$, $\6Z\gets \sigma(\rhat{\6H})$\;

Set $K_1 = \{n-r+\ell-w+1,\cdots,n-r+\ell-w_2\}$, $K_2 = \{n-r+\ell-w_2,\cdots,n-r+\ell\}$\;

Prepare 
$\mathcal K_1 = \left\{\left.\big(\6y_1,\6y_1\6Z_{K_1}^\top\big)\right|\6y_1\in\comb_{w_1}(\6c)\right\},$
$\mathcal K_2 = \left\{\left.\big(\6y_2,\rhat{\6s}-\6y_2\6Z_{K_2}^\top\big)\right|\6y_2\in\comb_{w_2}(\6c)\right\}$\;
$\mathcal K\gets \mathcal K_1\bowtie \mathcal K_2$\;
Compute $\6M\in\GL_{r,r}$ such that $\6M\sigma(\6H) = (\6U, \6I_r)$\;
$\rhat{\6s}\gets \6s\6M^\top$\;
$\widetilde{\6H}\gets$ matrix formed by rows and columns of $(\6U, \6I_r)$ at positions $\{d+1,\cdots,\ell\}$ and $\{1,\cdots,n-r+\ell\}$\; 
Set $L_1 = \{1,\cdots,n-r+\ell-w\}$ and $L_2 = \{n-r+\ell-w+1,\cdots,n-r+\ell\}$\;
Prepare $\mathcal L_1 = \left\{\left.\big(\6x_1,\6x_1\widetilde{\6H}_{L_1}^\top\big)\right|\6x_1\in\comb_{n-r+\ell-w}(\6c)\right\}$,
$\mathcal L_2 = \left\{\left.\big(\6x_2,\rhat{\6s}-\6x_2\widetilde{\6H}_{L_2}^\top\big)\right|\6x_2\in\mathcal K\right\}$\;
$\mathcal L \gets \mathcal L_1\bowtie \mathcal L_2$\;
\For{$\6x\in\mathcal L$}{
Plug $\6x$ into \eqref{eq:key_relation} to get $\widetilde{\6c}$\;
\SetAlgoNoLine
\If{$\widetilde{\6c}\in\comb_{n}(\6c)$}{
Return $\sigma^{-1}(\widetilde{\6c})$;
}
}
 \caption{New algorithm to solve PKP}
 \label{alg:novel}
\end{algorithm}

The correctness of the algorithm can be easily proven by considering that it essentially corresponds to the Algorithm 1 of \cite{pkpsolve2019}, plus an additional filtering stage in which we cut some of the candidates for $\mathcal L_2$.
To do this, we first find a $d$-dimensional subcode of $\code^\bot$, generated by $\rhat{\6H}$, with support size $w$.
We then find $\6S\in\GL_{d,r}$ such that $\rhat{\6H} = \6S\6H$ and compute $\rhat{\6s} = \6s\6S^\top$.
Recalling \eqref{eq:subcode_key_equation}, we use $\rhat{\6H}$ to produce candidates for the entries of $\widetilde{\6c}$ in the positions indexed by the support of $\sigma(\rhat{\6H})$, that is, $\{n-r+\ell-w+1,\cdots,n-r+\ell\}$.
To do this, we use a meet-in-the-middle approach (lines 4--6).
We then apply another transformation (lines 7--9) to both $\6H$ and $\6s$, employing the systematic form of $\6H$ to obtain $\ell-d$ new kernel equations involving exactly $n-r+\ell$ entries.
We then have another round of lists merging, corresponding to the same procedure employed in Algorithm 1 of \cite{pkpsolve2019}, with the only difference that to build $\mathcal L_2$ we use the elements of $\mathcal K$, instead of those in $\comb_{w}(\6c)$.
This difference is crucial since the gain of our method lies in this step: we expect $|\mathcal K|<|\hspace{0.5mm}\comb_{w}(\6c)|$, which yields a final list $\mathcal L$ with less elements.

In the following Proposition we derive the time complexity of the proposed algorithm.
\begin{proposition}\label{prop:new_time}
Let $d, w_1, w_2$ such that $\rcheck{N}_{w_1+w_2,d}>1$.
Then, Algorithm \ref{alg:novel} runs in time
$$T^{(d)}_{\scaleto{\mathcal{ISD}}{4pt}}(n,r,w_1, w_2)+T_{\mathcal K}+T_{\mathcal L}+\frac{n!q^{-\ell}}{(r-\ell)!},$$
with $w = w_1+w_2$ and $$T_{\mathcal K} = \frac{n!}{(n-w_1)!}+\frac{n!}{(n-w_2)!}+\frac{(n!)^2q^{-d}}{(n-w_1)!(n-w_2)!},$$ \begin{align*}
T_{\mathcal L} =  \frac{n!}{(r+w-\ell)!}& +\frac{n!q^{-d}}{(n-w)!}\\\nonumber
& \hspace{-0.5cm}+\frac{(n!)^2q^{-\ell}}{(n-w)!(r+w-\ell)!}.    
\end{align*}
\end{proposition}
\begin{IEEEproof}
Since $\rcheck{N}_{w,d}>1$, we expect $\code^\bot$ to contain at least a subcode with dimension $d$ and support size $w$.
To find such a subcode, we have a cost given by $T^{(d)}_{\scaleto{\mathcal{ISD}}{4pt}}(n,r,w)$.
Steps 2--4 come with a negligible cost, so we omit them.
The cost of building and merging the lists (using a smart binary search algorithm to determine the collisions) is given by $\mathcal K_1$ and $\mathcal K_2$ and results in $\frac{n!}{(n-w_1)!} + \frac{n!}{(n-w_2)!}$ operations.
The number of collisions, on average, is given by $|\mathcal K_1|\cdot |\mathcal K_2|\cdot q^{-d}$, so that the cost to produce $\mathcal K$ is
$$\frac{n!}{(n-w_1)!}+\frac{n!}{(n-w_2)!}+\frac{(n!)^2q^{-d}}{(n-w_1)!(n-w_2)!}.$$
After the collisions are checked, $\mathcal K$ contains $\frac{n!}{(n-w)!}q^{-d}$ elements, on average.
The cost of steps 7--9 can be neglected while, to execute steps 11-12, we repeat the previous reasoning and hence their cost is given by $|\mathcal L_1|+|\mathcal L_2|+|\mathcal L_1|\cdot|\mathcal L_2|\cdot q^{-(\ell-d)}$, which, on average, is equal to
\begin{align*}
\frac{n!}{(r+w-\ell)!}+\frac{n!q^{-d}}{(n-w)!}+\frac{(n!)^2q^{-\ell}}{(n-w)!(r+w-\ell)!}.
\end{align*}
Notice that, in the above formula, we have considered that $|\mathcal L_2| = |\mathcal K|$.
Finally, we also take into account the cost of iterating steps 13-16, whose number can be considered equal to the size of $\mathcal L$, and so, on average, is equal to $\frac{n!q^{-\ell}}{(r-\ell)!}$.
\end{IEEEproof}

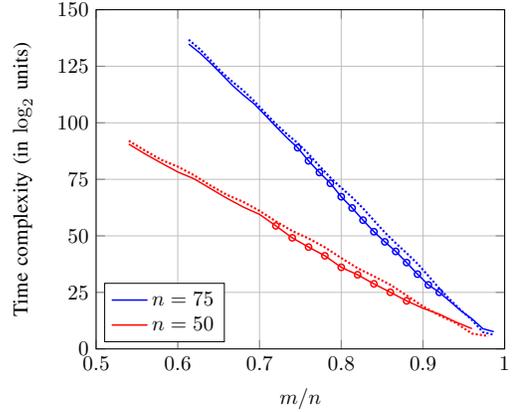
\begin{figure}
\centering
\resizebox{0.8\columnwidth}{!}{
\begin{tikzpicture}

\begin{axis}[
xtick={0, 0.2, 0.4, 0.5, 0.6, 0.7, 0.8, 0.9, 1},            
ytick={0,25,50,...,150},
xmin = 0.5,
xmax= 1,
grid = major,
ymin=0,
ymax=150,
legend style={at={(0.02, 0.02)},anchor=south west},
mark size=2pt,
xlabel={$m/n$},
ylabel={Time complexity (in $\log_2$ units)},
xticklabel style={
        /pgf/number format/fixed,
        /pgf/number format/precision=5
}]

\addplot[blue, line width=0.7pt]coordinates{
(0.613333333333333, 134.809526926436)
 (0.626666666666667, 130.982365437848)
 (0.640000000000000, 126.446638996092)
 (0.653333333333333, 121.602025561875)
 (0.666666666666667, 116.599180006746)
 (0.680000000000000, 112.308310924587)
 (0.693333333333333, 108.408502581467)
 (0.706666666666667, 103.442028467754)
 (0.720000000000000, 98.3760165547671)
 (0.733333333333333, 93.4472602746770)
 (0.746666666666667, 88.9861843837067)
 (0.760000000000000, 83.1830904826574)
 (0.773333333333333, 78.0289071186269)
 (0.786666666666667, 73.2569036379334)
 (0.800000000000000, 67.2981658361088)
 (0.813333333333333, 62.3322096075163)
 (0.826666666666667, 56.9549573150746)
 (0.840000000000000, 51.8262879385607)
 (0.853333333333333, 47.3459215719210)
 (0.866666666666667, 43.1106744033118)
 (0.880000000000000, 38.1817552480058)
 (0.893333333333333, 33.0628973062025)
 (0.906666666666667, 28.3105012963788)
 (0.920000000000000, 25.0277405348958)
 (0.933333333333333, 20.9647070508082)
 (0.946666666666667, 16.8175388330135)
 (0.960000000000000, 13.3043113712480)
 (0.973333333333333, 8.93293663385080)
 (0.986666666666667, 7.61562828731388)
};\addlegendentry{$n = 75$}

\addplot[red, line width=0.7pt]coordinates{
(0.540000000000000, 90.5893240373760)
 (0.560000000000000, 86.2176960983173)
 (0.580000000000000, 82.1729909422834)
 (0.600000000000000, 78.2536545193600)
 (0.620000000000000, 75.2295708489656)
 (0.640000000000000, 70.9138991569144)
 (0.660000000000000, 66.6525764480577)
 (0.680000000000000, 62.7590741449817)
 (0.700000000000000, 59.4978102351293)
 (0.720000000000000, 54.4264181060127)
 (0.740000000000000, 49.1342912333412)
 (0.760000000000000, 45.0407357397065)
 (0.780000000000000, 41.1690592323635)
 (0.800000000000000, 36.0806405812159)
 (0.820000000000000, 32.7888870114744)
 (0.840000000000000, 28.7689249852703)
 (0.860000000000000, 25.0494960544378)
 (0.880000000000000, 21.2502840740398)
 (0.900000000000000, 17.9815251079667)
 (0.920000000000000, 15.3747470235311)
 (0.940000000000000, 11.9141684217708)
 (0.960000000000000, 8.96632098289655)
}; \addlegendentry{$n = 50$};

\addplot[red, densely dotted,line width=1pt]coordinates{
(0.540000000000000, 92.0294860472408)
(0.560000000000000, 87.6841578336334)
(0.580000000000000, 83.5956874787257)
(0.600000000000000, 80.6169331018905)
(0.620000000000000, 76.7790660481559)
(0.640000000000000, 72.2598365921026)
(0.660000000000000, 68.1651550846510)
(0.680000000000000, 65.0578642977148)
(0.700000000000000, 61.0660407293382)
(0.720000000000000, 56.4242214273558)
(0.740000000000000, 52.1861744364421)
(0.760000000000000, 48.8677726492863)
(0.780000000000000, 44.9114093505454)
(0.800000000000000, 40.1979543266853)
(0.820000000000000, 35.7011746898132)
(0.840000000000000, 32.0978715360616)
(0.860000000000000, 28.3771907159599)
(0.880000000000000, 23.6304893569071)
(0.900000000000000, 18.7946078791289)
(0.920000000000000, 14.8144994859996)
(0.940000000000000, 11.5449875160946)
(0.960000000000000, 6.78083709260212)
(0.980000000000000, 5.67804945263712)
};

\addplot[red, only marks, mark = o, mark size=1.5pt,line width=0.7pt]coordinates{
(0.720000000000000, 54.4264181060127) (0.740000000000000, 49.1342912333412)
 (0.760000000000000, 45.0407357397065)
 (0.780000000000000, 41.1690592323635)
 (0.800000000000000, 36.0806405812159)
 (0.820000000000000, 32.7888870114744)
 (0.840000000000000, 28.7689249852703)
 (0.860000000000000, 25.0494960544378)
 (0.880000000000000, 21.2502840740398)
};

%%%%%%%%%%%%%%%%%%%%%%%%%%%%%%%%%%

\addplot[blue, densely dotted,line width=1pt]coordinates{
(0.613333333333333, 136.757924752043)
 (0.626666666666667, 132.629699540759)
 (0.640000000000000, 127.619974882112)
 (0.653333333333333, 122.547244229020)
 (0.666666666666667, 118.189127699753)
 (0.680000000000000, 114.273313040022)
 (0.693333333333333, 109.672406916160)
 (0.706666666666667, 104.475818534957)
 (0.720000000000000, 99.4562193393849)
 (0.733333333333333, 95.0721697079172)
 (0.746666666666667, 91.0167302760961)
 (0.760000000000000, 86.1936886655368)
 (0.773333333333333, 80.8656509470683)
 (0.786666666666667, 75.7745314904393)
 (0.800000000000000, 71.2549837954978)
 (0.813333333333333, 67.0252234006945)
 (0.826666666666667, 62.1836816662235)
 (0.840000000000000, 56.7859888884923)
 (0.853333333333333, 51.5119244990118)
 (0.866666666666667, 46.7754823559122)
 (0.880000000000000, 42.3448365905593)
 (0.893333333333333, 37.6963191945558)
 (0.906666666666667, 32.2822279100211)
 (0.920000000000000, 26.7408949214623)
 (0.933333333333333, 21.7037747583249)
 (0.946666666666667, 17.0299247743908)
 (0.960000000000000, 12.8303897349568)
 (0.973333333333333, 7.42970264519059)
 (0.986666666666667, 6.25358855233763)
};

\addplot[blue, only marks, mark = o, mark size=1.5pt,line width=0.7pt]coordinates{
 (0.746666666666667, 88.9861843837067)
 (0.760000000000000, 83.1830904826574)
 (0.773333333333333, 78.0289071186269)
 (0.786666666666667, 73.2569036379334)
 (0.800000000000000, 67.2981658361088)
 (0.813333333333333, 62.3322096075163)
 (0.826666666666667, 56.9549573150746)
 (0.840000000000000, 51.8262879385607)
 (0.853333333333333, 47.3459215719210)
 (0.866666666666667, 43.1106744033118)
 (0.880000000000000, 38.1817552480058)
 (0.893333333333333, 33.0628973062025)
 (0.906666666666667, 28.3105012963788)
 (0.920000000000000, 25.0277405348958)
};

\end{axis}
\end{tikzpicture}
}
\caption{Comparison of the time complexity of \cite[Algorithm 1]{pkpsolve2019} (dotted lines) with that of our algorithm (full lines), for $q = 251$. The circles highlight the cases in which our algorithm is optimized with $d\geq 2$.}
\label{fig:n100}
\end{figure}

In Figure \ref{fig:n100} we compare the performance of Algorithm \ref{alg:novel} with that of \cite[Algorithm 1]{pkpsolve2019}, for the case of $q = 251$ and several pairs of values $(m,n)$, chosen such that $n!q^{-m}<1$.
As we can see, unless $m$ is close to $n$, our algorithm is faster than the one in \cite{pkpsolve2019}; in particular, the speed-up increases when our algorithm is optimized with $d\geq 1$. 

To assess the impact of our algorithm on the cryptanalysis of schemes relying on the PKP, in Table \ref{tab:attack} we consider the PKP-DSS instances which have been recommended in \cite{beullens2019pkp} for the security levels of 128 and 192 bits. For these instances, the claimed cost of \cite[Algorithm 1]{pkpsolve2019} is $2^{130}$ and $2^{193}$, respectively.
As we can see, our attack is faster and, furthermore, has a cost which is slightly lower than the claimed security levels.
For the 256-bit security instance, we found instead that our attack does not improve upon \cite{pkpsolve2019}. 
\begin{table}[t]
\centering
\caption{Time complexity of our attack for the PKP-DSS parameters recommended in \cite{pkpsolve2019, beullens2019pkp}.}
\begin{tabular}{|c|c|cc|}\hline
$(n,m,q)$   & Claimed cost & $(d,w,w_1,w_2, \ell)$ & Cost of Algorithm \ref{alg:novel}  \\\hline
$(69,41,251)$ & $2^{130}$ & $(1,22, 2, 20, 16)$ & $2^{125.47}$    \\
$(94, 54, 509)$ & $2^{193}$ & $(1, 31, 2, 29, 22)$ & $2^{189.77}$    \\\hline
\end{tabular}
\label{tab:attack}
\end{table}

\section{Conclusion}\label{sec:conclusions}

We have described a novel attack to the PKP which makes use of small support subspaces of kernel equations.
Our proposed algorithm is based on techniques borrowed from the code-based cryptography context and is faster than state-of-the-art attacks for several cases. 
To consider a situation of practical interest, we have shown that the security of some PKP-DSS instances is slightly overestimated.
Despite the moderate gain in complexity with respect to the state-of-the-art, our work shows that the PKP can be solved exploiting coding theory techniques and this may lead to new, possibly even more efficient, attack avenues in the future.
\bibliographystyle{IEEEtran}
\bibliography{References}

\end{document}